\documentclass[%
 reprint,
 amsmath,amssymb,
 aps,
]{revtex4-2}

\usepackage{xcolor}  
\usepackage{graphicx}
\usepackage{dcolumn}
\usepackage{bm}

\begin{document}

\preprint{APS/123-QED}

\title{Electric-field control of zero-dimensional topological states in ultranarrow germanene nanoribbons}

\author{Lumen Eek\textsuperscript{1}}
\thanks{These authors contributed equally to this work.}
\author{Esra D. van 't Westende\textsuperscript{2}}%
\thanks{These authors contributed equally to this work.}

\author{Dennis J. Klaassen\textsuperscript{2}}%
\author{Harold J. W. Zandvliet\textsuperscript{2}}%
\author{Pantelis Bampoulis\textsuperscript{2}}%
\email{p.bampoulis@utwente.nl}
\author{Cristiane Morais Smith\textsuperscript{1}}
\email{c.demoraissmith@uu.nl}

\affiliation{\textsuperscript{1}Institute for Theoretical Physics, Utrecht University, Princetonplein 5, 3584 CC Utrecht, the Netherlands}
\affiliation{\textsuperscript{2}Physics of Interfaces and Nanomaterials, MESA+ Institute, University of Twente, Drienerlolaan 5, 7522 NB Enschede, the Netherlands}

\date{\today}

\begin{abstract}
Reversible, all-electric control of symmetry-protected zero-dimensional modes has been a long-standing goal. In buckled honeycomb lattices, a perpendicular field couples to the staggered sublattice potential providing the required handle. We combine scanning tunneling microscopy and tight-binding theory to switch zero-dimensional topological end states reversibly on and off in ultranarrow germanene nanoribbons by tuning the electric field in the tunnel junction. Increasing the field switches off the end modes of topological two-hexagon wide ribbons, while the same field switches on zero-dimensional states in initially trivial three- and four-hexagon wide ribbons. This atomic scale platform realizes a proof-of-principle for a zero-dimensional topological field effect device, opening a path for ultrasmall memory, controllable qubits, and neuromorphic architectures.
\end{abstract}

\maketitle

\textit{Introduction.~}Topological insulators (TIs) have transformed condensed-matter physics by revealing the existence of symmetry-protected boundary states that are beyond conventional band theory descriptions \cite{hasan2010colloquium, kane2005z2, kane2005quantum, bernevig2006quantum, chiu2016classification}. In particular, zero-dimensional (0D) modes have attracted much attention because of their potential use as qubits in quantum computers \cite{kitaev2001unpaired, alicea2012new, sarma2015majorana}. Experimentally, 0D states have been observed as corner modes in higher-order TIs \cite{kempkes2019robust, schindler2018higher, canyellas2024topological} and localized states in topological superconductors \cite{qi2011topological, beenakker2013search, jack2019observation, alicea2012new}. In 1D honeycomb systems, 0D topological states have emerged as boundary states in staggered-width and chiral graphene nanoribbons \cite{rizzo2018topological, rizzo2018topological, li2021topological}, and germanene nanoribbons \cite{klaassen2025realization}. For device applications, these 0D modes must be reversibly controllable by an external knob, ideally using an electric field, which is compatible with nanoscale gating \cite{gilbert2021topological, collins2018electric, bampoulis2023quantum, qian2014quantum, matthes2014influence, ezawa2013quantized}. The application of a perpendicular electric field in buckled honeycomb lattices couples directly to the sublattice potential \cite{drummond2012electrically, ezawa2013quantized, Ezawa2015ML}, and was shown to drive a topological phase transition in 2D TIs such as germanene and Na$_3$Bi \cite{collins2018electric, bampoulis2023quantum}. However, no experiment has achieved electric field controlled switching of 0D topological modes, and the experimental realisation of a 0D topological field-effect device remains open.

Here we demonstrate reversible, electric field switching of 0D topological end modes in epitaxially grown germanene nanoribbons, that are only two to four hexagons wide ($\sim$1-2 nm). Using 77 K scanning tunnelling microscopy/spectroscopy (STM/STS) and tight-binding calculations, we tune the STM vertical tunnel junction electric field to (i) switch off the end states of intrinsic 1D TIs (two-hexagon wide ribbons) and (ii) switch on end states and thus induce topology in ribbons that are trivial at zero field (three- and four-hexagon wide ribbons). The switching is reversible and repeatable. These results establish the smallest-possible 0D topological field-effect device and close the outstanding gap between static observation and dynamic control of 0D modes.

\begin{figure*}[t]
  \centering
  \includegraphics[width=0.9\textwidth]{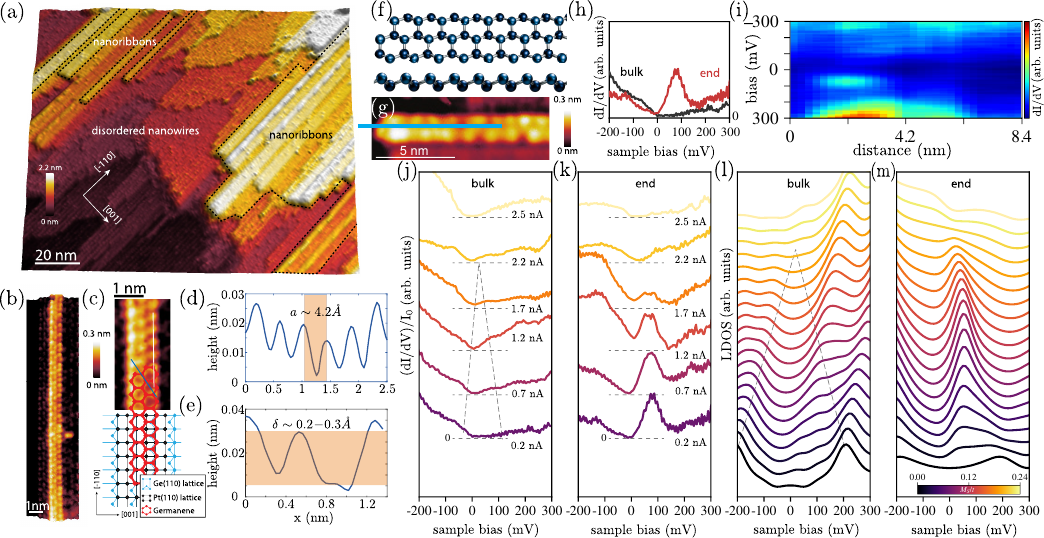}
  \caption{(a) Large area STM topograph showing germanene nanoribbons (black dashed outlines) embedded in disordered nanowires on the Pt/Ge(110) surface. The ribbons run along the [-110] direction. (b) High resolution image of a two-hexagon wide nanoribbon. (c) Atomic resolution image of the two-hexagon nanoribbon overlaid on the tentative structural model. (d) and (e) Line profiles taken along the white and teal lines in panel (c), respectively, showing the lattice periodicity and buckling in the hexagonal lattice. (f) Schematic of the buckled honeycomb lattice. (g) STM image of a narrow nanoribbon; the cyan line indicates the spatial path of the line spectroscopy. (h) $dI(V)/dV$ spectra recorded at the end (red) and bulk (black) of the nanoribbon in (g). (i) $dI(V)/dV$ line spectroscopy recorded along the cyan line in (g), showing  the localized end state. (j,k) $(dI(V)/dV)/I_0$ spectra recorded for increasing current setpoints ($I_0$) from I=0.2 nA to I=2.5 nA, for the bulk (j) and end (k) of the nanoribbon in (g). The images have been displaced for clarity; zero is represented by a dashed line. (l,m) Theoretically calculated LDOS, for several values of the staggered mass $M_S$, see inset in (m) for the color code. The results for the bulk are shown in (l) and for the edge in (m). The plots obtained for the lowest values of staggered mass, black curves in (l) and (m), should not be compared to the experiments because germanene always has a finite mass due to the buckling. The dashed lines in (j) and (l) indicate the closing of the bulk gap. In the corresponding regime of parameters, a (slightly displaced) zero-bias peak is observed at the ends, see (k) and (m).}
  \label{fig:fig1}
\end{figure*}

Germanene is a low-buckled honeycomb monolayer of Ge atoms \cite{cahangirov2009two, bampoulis2014germanene,zhang2016structural, bampoulis2023quantum, acun2015germanene, molle2017buckled, reis2017bismuthene, bechstedt2021beyond} that hosts a spin-orbit coupling (SOC) induced gap at the K/K' points of the Brillouin zone and, in its 2D form, it is a Kane-Mele-type TI (class AII) with helical edge states \cite{matthes2014influence, Ezawa2015ML, liu2011quantum, bampoulis2023quantum, zandvliet2024evidence}. When narrowed below $\sim$2.5 nm, germanene nanoribbons undergo a transition into a 1D topological insulator with in gap end modes protected by time reversal and mirror symmetries \cite{klaassen2025realization}. We have grown ultranarrow zigzag terminated germanene nanoribbons using a segregation based epitaxy on Pt/Ge(110), as described in Refs.~\cite{yuhara2021epitaxial, klaassen2025realization} (see also methods for details). Figure~\ref{fig:fig1}(a) provides an STM overview of epitaxial germanene nanoribbons on Pt/Ge(110). The nanoribbons run along the [-110] direction and reside within a disordered nanowire background produced during growth. The left-hand side of the image contains isolated nanoribbons, whereas the right-hand side shows a densely packed nanoribbon domain. The nanoribbons exhibit varying widths (typically $\sim$1-5 nm) that arise stochastically during growth and are not controllable at this stage.

A high-resolution STM image of a two-hexagon wide nanoribbon is shown in Fig.~\ref{fig:fig1}(b). Its atomic scale lattice is revealed in Fig.~\ref{fig:fig1}(c), where the overlaid tentative ball and stick model \cite{klaassen2025realization} depicts a low-buckled honeycomb geometry. From the line profiles in Fig.~\ref{fig:fig1}(d) and Fig.~\ref{fig:fig1}(e), we extract a lattice constant of about 4.2 \AA{} (often a larger centered rectangular periodicity is also visible \cite{klaassen2025realization}) and a buckling of 0.2-0.3 \AA{} for this ultranarrow nanoribbon. Owing to the symmetry mismatch with the Pt/Ge(110) surface, the ribbon couples only weakly to the substrate and terminates in zigzag edges along the [-110] direction \cite{klaassen2025realization}. The ribbons are morphologically and electronically distinct from both the Pt/Ge substrate and surrounding nanowires, see Fig. S2. Ribbons wider than six hexagons ($\sim$2.5 nm) are 2D TIs with helical 1D edge states, whereas narrower ribbons transition to a 1D topological insulator with 0D end modes \cite{klaassen2025realization}. The appearance of these end states depends on the ribbon width, intrinsic spin-orbit coupling, Semenoff mass, and higher-order hopping parameters in a Kane-Mele type model \cite{klaassen2025realization, traverso2024emerging}. Nevertheless, the end states, protected by a quantized Zak phase, were proven to exist in a broad regime of parameters, to be experimentally accessible with STM, and to be robust \cite{klaassen2025realization}.

\begin{figure*}[t]
  \centering
  \includegraphics[width=0.9\textwidth]{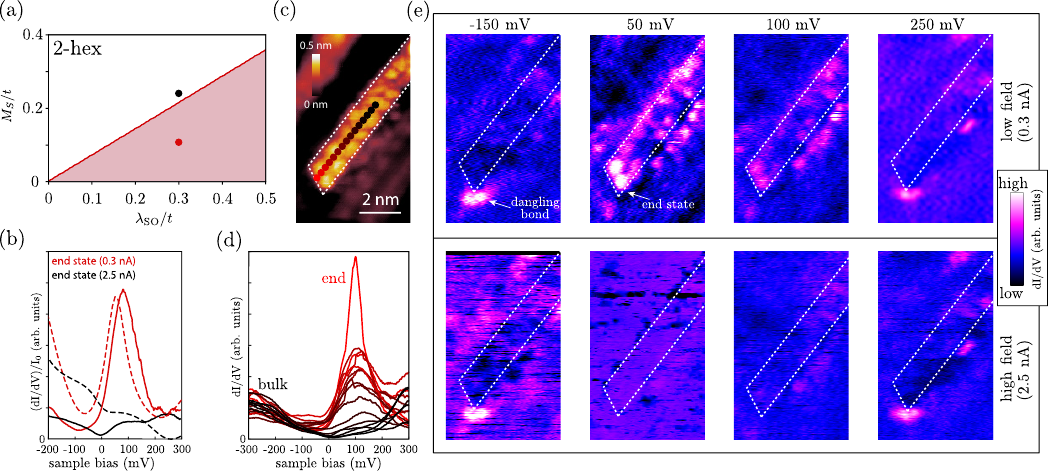}
  \caption{(a) Topological phase diagram in the $\lambda_{\text{SO}}-M_{\text{S}}$ plane for two-hexagon wide nanoribbons. The topological regime is depicted in red and the trivial regime in white. (b) Experimental $(dI(V)/dV)/I_0$ (solid lines) and calculated LDOS (dashed lines of the end state at low (red) and high (black) electric fields. (c) STM topography of an ultranarrow nanoribbon (two-hexagon wide), on which the spectra in (b) are recorded. (d) $dI(V)/dV$ point spectra acquired moving from the end ribbon (red) to the bulk (black) of the ribbon (see colored points in c for the location, the color of the marker corresponds to the color of the spectrum), showing the sharp decay of the localized end state.  (e) dI/dV maps recorded at bias voltages of -150, 50, 100 and 250 mV under low (0.3 nA) and high (2.5 nA) current setpoints, illustrating the localization of the end state (50-100 mV) at the ribbon end and its disappearance at high electric fields. In contrast, the dangling bonds are still visible at high fields.}
  \label{fig:fig2}
\end{figure*}

The existence of robust 0D end modes in an intrinsic low-buckled honeycomb geometry makes ultranarrow germanene ribbons an ideal platform to explore electric field-driven topological switching. The out of plane buckling, see cartoon in Fig.~\ref{fig:fig1}(f), separates the inverted $\pi$ orbitals onto distinct sublattices, which is crucial for field induced phase transitions by breaking the band-inversion symmetry. We examined the effect of a perpendicular electric field on the electronic structure of these ultranarrow nanoribbons, and particularly on the end states. This was achieved by leveraging the built-in electric field in the STM tunnel junction. The $\sim$1.5 eV work-function offset between a PtIr tip ($\sim$5.5 eV) and germanene ($\sim$4 eV \cite{borca2020image}) produces an electrostatic potential that scales with the tip-sample separation. This separation is set with sub-\AA{} precision through the tunnelling-current setpoint $I_0$. Higher $I_0$ results into a smaller tip-sample distance and thus a stronger electric field \cite{collins2018electric,bampoulis2023quantum}. Because our sample voltage bias is small in comparison to the electrostatic potential difference, it can be neglected. Note however that absolute field estimates are uncertain owing to unknown tip geometry, image charges, and possible band bending \cite{collins2018electric,bampoulis2023quantum}, but the relative change is exactly given by $I_0$. We therefore report all field dependencies in terms of current; low fields correspond to $I_0 = 0.2-0.5$ nA, and high fields to $I_0 = 2-3$ nA. 

\begin{figure*}[t]
  \centering
  \includegraphics[width=0.9\textwidth]{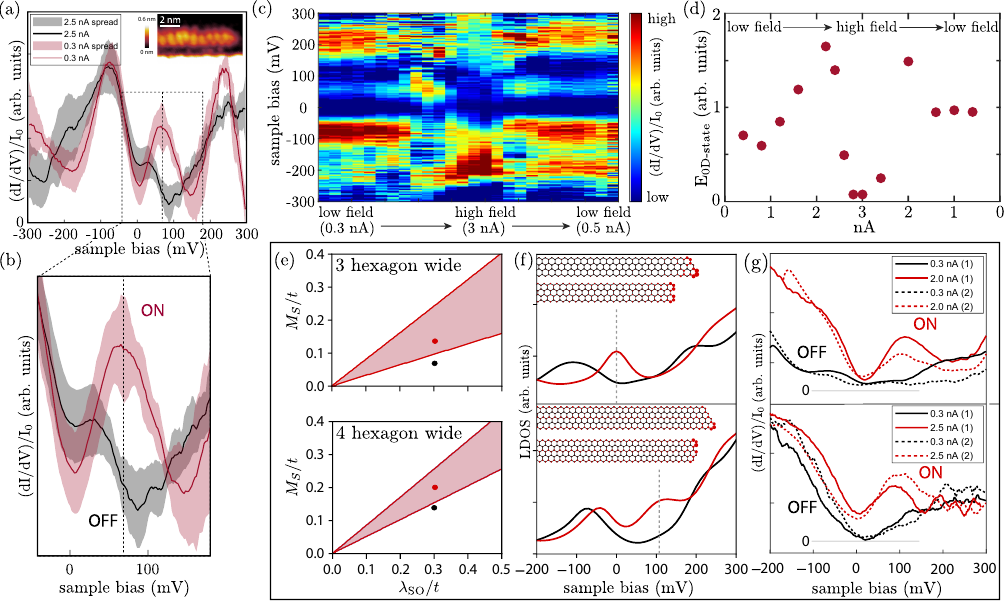}
  \caption{(a) Repeated cycling (seven cycles) between low ($0.3\,$nA) and high ($2.5\,$nA) applied perpendicular electric fields  at the end of a two-hexagon ribbon, demonstrating reproducible and reversible on/off switching of the end states. The inset shows the STM topography of the nanoribbon used to perform the cycling experiments. (b) Zoom-in of the dashed-box region in (a), highlighting the discrete switching of the 0D end state. (c) $(dI(V)/dV)/I_0$ and (d) the normalized intensity of the end mode as a function of a gradual change of the setpoint current, sweeping it from low to high and back. (e) Topological phase diagrams calculated for different values of $M_\text{S}$ and $\lambda_{\text{SO}}$ for the three- (top) and four- (bottom) hexagon wide nanoribbons. The topological regime is depicted in red and the trivial regime in white. (f) Calculated LDOS of the three- and four-hexagon wide nanoribbons, showing the predicted end states (localized at the end of the nanoribbons, see insets) at large perpendicular electric fields and their absence at small electric fields. (g) Experimental $(dI(V)/dV)/I_0$ spectra for three-hexagon (top) and four-hexagon (bottom) ribbons at low and high setpoints, showing reversible end-state appearance around $50$-$100\,$mV at high electric fields. Two cycles of low-high-low-high electric fields are depicted for both ribbons.
}
  \label{fig:fig3}
\end{figure*}

An STM topograph of a $\sim$ 1 nm wide (two hexagon) germanene nanoribbon is shown in Fig.~\ref{fig:fig1}(g). The $dI/dV$ spectra, recorded at the ribbon end (red) and interior/bulk (black), are presented in Fig.~\ref{fig:fig1}(h). A sharp end state peak marks the 0D topological mode \cite{klaassen2025realization}, whereas the bulk spectrum displays a conductance minimum that we assign to the bulk gap. This conductance minimum does not reach zero because substrate contributions and thermal broadening offsets and convolutes the entire spectrum, and does not allow to resolve the minigap. These effects limit us from quantitative determining the gap size. Figure~\ref{fig:fig1}(i) depicts the corresponding $dI(V)/dV$ line spectroscopy as we move from the end of the nanoribbon to about 8 nm inwards (bulk), confirming that the density of states (DOS) peak is localized at the ribbon end and decays rapidly into the bulk. The response of the nanoribbon to the electric field (tuned via the current setpoint) is presented in Figs.~\ref{fig:fig1}(j)-(m). Figures~\ref{fig:fig1}(j) and (k) show the $(dI(V)/dV)/I_0$, i.e. the differential conductance normalized to the current setpoint $I_0$. This yields internally consistent comparisons of line shapes and relative intensities at different setpoints, see End Matter for details on the normalization. Increasing the current set point and hence the vertical electric field progressively suppresses the end state, which completely dissapears around 2.2 nA, see Fig.~\ref{fig:fig1}(k). The $(dI(V)/dV)/I_0$ spectra of the bulk, Fig.~\ref{fig:fig1}(j), mirrors this evolution. As the current set point is increased, the bulk gap-like feature narrows, closes at 1.7-2.2 nA, and then reopens as a trivial gap (the conductance minimum broadens, and there are no end states). The behavior of the bulk confirms the bulk-boundary correspondence. Trivial substrate states and dangling bonds are unaffected, see Supplemental Materials (SM) \footnote{See Supplemental material at XXX, which includes Refs.~\cite{zhang2015probing, murray2019comprehensive, Ezawa2015ML, klaassen2025realization, traverso2024emerging, SSH, rizzo2018topological}, for additional data on the influence of electric field on trivial surface states; additional calculations of the topological invariants of germanene nanoribbons; and an example of a 0D state pinning at the transition of a 2-hexagon wide to 3-hexagon wide ribbon.}.

These results are corroborated by theoretical calculations using a tight-binding description based on the Kane-Mele model for graphene \cite{TBGermanene, kane2005quantum}. In addition, longer-range hopping must be taken into account because first-principles calculations have indicated that the next-next-nearest neighbor hopping parameter is sizable, and even larger than the next-nearest neighbor one \cite{klaassen2025realization, traverso2024emerging, reich2002tight}, see End Matter. The calculated local density of states (LDOS) spectra for the bulk and end are presented in Figs.~\ref{fig:fig1}(l) and (m), respectively. Additionally, we present band structures in Figs.~S8-S9 of the SM. The main difference with graphene is that germanene has a large intrinsic SOC (with a SOC gap estimated to be of the order of 100 meV when germanene is in proximity to Pt \cite{bampoulis2023quantum}) and a staggered mass M=10-50 meV, which must be included because of the inherently buckled structure of germanene. This also implies that the lowest plots of the LDOS for nearly vanishing staggered mass [black lines in Figs.~\ref{fig:fig1}(l) and (m)] should not be compared to the experiments because this mass is never zero. We nevertheless included them for completeness. The agreement between theory and experiments is remarkable. This behavior can be understood by recalling that the electric field breaks inversion symmetry and induces a staggered sublattice potential (a tunable Semenoff mass, $M_{\text{S}}$), which through a complex interplay with the intrinsic SOC, $\lambda_{\text{SO}}$, leads to a rich phase diagram. For the two-hexagon wide ribbon, the topological gap closes in a continuous way at a critical electric field. Upon further increasing the field, the gap reopens with a trivial Zak phase, and the 0D end modes are no longer present. We have confirmed this bulk-boundary correspondence via the comparison between the closure/reopening of the gap with both, the calculated Zak phase and the STM spectra (see SM).

The topological phase diagram in the $\lambda_{\text{SO}}-M_{\text{S}}$ plane is shown in Fig.~\ref{fig:fig2}(a) for the two-hexagon wide ribbon. For $M_{\text{S}}<\lambda_{\text{SO}}$ (red area) the ribbon is in the 1D TI phase, while larger staggered mass drives it into a trivial insulating phase (white area). The red and black dots mark the theoretical parameters that match with the experimental data under the low and high field STM conditions, respectively, highlighting that the applied electric field is strong enough to push the system across the topological phase boundary. Figure~\ref{fig:fig2} provides additional evidence and real space $dI/dV$ maps of this topological phase transition. Figure ~\ref{fig:fig2}(b) shows $(dI(V)/dV)/I_0$ recorded at the end of the $\sim$1 nm wide ribbon in Fig.~\ref{fig:fig2}(c), and the calculated LDOS spectra of the end state at low (red) and high (black) electric fields. There is an excellent agreement between theory and experiment. The blue arrow in panel (c) indicates the line along which $(dI(V)/dV)$ spectra in Fig.~\ref{fig:fig2}(d) were taken. By moving from the ribbon bulk (black) to the ribbon end (red) a sharp in gap state centered around +100 mV emerges, providing another example of the exponentially localized end mode that is controlled by the tip-induced electric field. The spatial conductance maps (spectral slices at the specified energy) in Fig.~\ref{fig:fig2}(e) confirm this behavior. Under the low field setpoint (top row), the end state is prominent between 50 mV and 100 mV, and confined to the ribbon end. When the same biases are probed at high field (bottom row) the end state disappears, whereas the signal from the dangling bonds (seen at -150 and 250 meV) persists. This confirms that the electric field selectively suppresses the topological end mode rather than globally perturbing the electronic structure. This behavior is also further discussed in the SM.

Figs.~\ref{fig:fig3}(a) and (b) demonstrate that the electric-field transition is reversible. For a two-hexagon ribbon the $(dI(V)/dV)/I_0$ spectra show that increasing the $I_0$ kills the 0D end modes (a topological to trivial phase transition), whereas decreasing $I_0$ restores the low field condition and the end modes re-emerge. The zoom in Fig.~\ref{fig:fig3}(b) highlights the discrete switching. Seven successive low to high to low electric field cycles reproduce the binary on/off behaviour with minor variations. Although the precise on/off thresholds and the amplitude of the recovered state vary from ribbon to ribbon and from cycle to cycle, probably due to slight tip changes or intrinsic variations in the phase boundary, the underlying electric field control is clearly demonstrated. The bidirectionality of this behavior is evident in Figs.~\ref{fig:fig3}(c) and (d), which track, respectively, the $(dI(V)/dV)/I_0$ and end mode intensity during gradual increase and subsequent gradual decrease of the $I_0$. It is interesting that the 0D mode first becomes more intense upon increasing the electric field, before disappearing, see Figs.~\ref{fig:fig3}(c) and (d), in agreement with theoretical calculations in Fig.~\ref{fig:fig1}(m).

Finally, we show that a perpendicular electric field can not only erase but also create topological order in germanene nanoribbons. In Figs.~\ref{fig:fig3}(e-g), we highlight the electric field effect on nanoribbons of different widths. In contrast to two-hexagon ribbons, three- and four-hexagon ribbons are topological for a smaller parameter space, see Fig.~\ref{fig:fig3}(e) and compare with Fig.~\ref{fig:fig2}(a). In practice, these wider ribbons are trivial at low electric fields (staggered mass). As the field increases, the trivial gap closes and then reopens with a non-zero Zak phase. This should lead to the emergence of topological end modes that are localized at the nanoribbon end (for both the three- and the four-hexagon nanoribbons), as shown in the calculations in Fig.~\ref{fig:fig3}(f) and their insets. Notice that the LDOS curves correspond to the top ribbon in the insets. We additionally display an end state on ribbons with a different termination. The black and red dots in Figs.~\ref{fig:fig3}(e) indicate the model parameters. This is confirmed experimentally in Figs.~\ref{fig:fig3}(g), where end states appear for both three- and four-hexagon ribbons at high electric fields (the ribbons are devoid of end states at low fields). Reversing the sweep restores the initial spectra, dotted lines in Fig.~\ref{fig:fig3}(g), demonstrating a fully reversible, electric field-driven trivial to topological transition. In the SM, we provide additional examples of this transition.

The reversible electric field switching of 0D end modes in ultranarrow ($\sim$ 1 nm wide) and ultrathin (one-atom thick) germanene nanoribbons presented here provides the first proof of principle for the realization of a 0D topological field effect device. Using the local gate formed by the STM tunnel junction, we achieve continuous closing and reopening of the bulk gap, while tracking the emergence and vanishing of the end states. Three independent experimental signatures, consistent with the Zak phase, establish that the end modes are topological: (i) the bulk-boundary correspondence of the 0D modes at the ribbon end and at domain walls; (ii) a one-to-one correspondence between gap closing/reopening and disappearance/appearance of end states under a controlled vertical electric field; (iii) and selectivity of the field response to the end mode only, while trivial features remain invariant. The experimentally achievable critical field (found to occur for setpoint currents in the range of $2-3$ nA and estimated to be around $2-2.5$ V/nm) and the minor variations confirm the robustness and reversibility of this mechanism. This atomic scale platform not only validates the concept of a topological switch at the smallest conceivable device footprint, but also opens immediate pathways toward ultra small non-volatile memory cells, electrically controllable topological qubits, and neuromorphic synapses based on symmetry-protected states.

\textit{Acknowledgements.} L.E., H.J.W.Z., and C.M.S. acknowledge the research program “Materials for the Quantum Age” (QuMat) for financial support. This program (registration number 024.005.006) is part of the Gravitation program financed by the Dutch Ministry of Education, Culture and Science (OCW). H.J.W.Z. and D.J.K are supported by NWO Grant OCENW.M20.232. P.B. acknowledges financial support from the European Research Council; funded by the European Union (ERC, Q-EDGE, 101162852). Views and opinions expressed are however those of the author(s) only and do not necessarily reflect those of the European Union or the European Research Council. Neither the European Union nor the granting authority can be held responsible for them.

\textit{Data availability.} The data that support the findings of this article are openly available at \footnote{L. Eek \textit{et al.}, Data for Electric-field control of zero-dimensional topological states in ultranarrow germanene nanoribbons, 10.4121/bd047595-feab-406e-85c3-c333a3b4dcc7}.

\bibliography{bibliography}

\section*{End matter}

\textit{Scanning tunneling microscopy.} STM data were acquired using a UHV Omicron low-temperature STM operating at 77 K. Samples were transferred from the preparation chamber, where they were grown, to the STM chamber under ultra-high vacuum conditions. The tips were made from PtIr chemically etched wires. Constant current STM imaging was performed with current setpoints typically ranging from 0.2 to 2.5 nA and voltage biases typically between -0.5 V to 0.5 V. The $dI(V)/dV$ spectra for STS were recorded with the feedback loop disabled, employing a lock-in amplifier with an AC modulation voltage of 10-20 mV at a frequency range of 1000-1200 Hz. Data processing and analysis of STM images and $dI(V)/dV$ spectra were carried out using MATLAB, Gwyddion, and SPIP. The $dI(V)/dV$ point and line spectra in our manuscript were taken in constant-height STS (feedback loop switched off after setting I$_0$,V$_{set}$), whereas the $dI/dV$ maps were acquired in constant-current STS (feedback loop switched on). Both modes are standard practice, and it is important when measuring $dI/dV$ maps to have the feedback loop on, otherwise the tip can crash onto the surface, since it is scanning the surface while measuring. This is not the case in point spectra, for which the tip remains stationary at a certain location. Now, because the two modes operate with different control parameters, their apparent amplitudes at a given bias do not necessarily coincide \cite{murray2019comprehensive}. 

The $dI(V)/dV$ spectra recorded at different current setpoints ($I_0$) require a normalization process to be directly compared, which was done by dividing by $I_0$. Furthermore, often (possibly) tip LDOS or band bending effects lead to an asymetric $dI(V)/dV$ spectrum. This asymetry changes with $I_0$ and therefore needs to be corrected before comparing the $(dI(V)/dV)/I_0$ spectra. The workflow for background correction and normalization of a single $dI(V)/dV$ spectrum is depicted in Fig. S3.  The raw $dI(V)/dV$ spectra are first divided by $I_0$ to normalize the amplitudes across different datasets recorded at different $I_0$. Thereafter, the normalized $(dI(V)/dV)/I_0$ spectra are corrected by subtracting a second-order polynomial in order to remove the slowly varying background that differs between different current setpoints and tips. After this correction, the two curves recorded at different current setpoints show good overlap at bulk states and the end states are better defined. This yields internally consistent comparisons of line shapes and relative intensities.\\

\textit{Nanoribbon growth.} Germanene nanoribbons were grown on Ge(110) in ultra-high vacuum (base pressure in low  10$^{-10}$ mbar) by first cleaning Ge(110) wafers via Ar$^+$ sputtering and high-temperature annealing (1150 K), then depositing about 1 ML of Pt and then flash annealing to 1150 K. Upon cooling down to room temperature, Ge atoms segregate on the Pt layer, forming extended, flat germanene ribbons when the underlying Pt coverage is near a full monolayer \cite{klaassen2025realization}. Lower local Pt coverage instead yields Ge nanowires \cite{klaassen2025realization}. Both ribbons and wires align along [-110] due to the substrate symmetry, but only the ribbons reach lengths of hundreds of nanometers with atomically uniform honeycomb structure, such as the ones in Fig.~\ref{fig:fig1}.\\

\textit{Theoretical model of 0D end states in ultranarrow germanene nanorribons.} Germanene is composed of a monolayer of germanium atoms arranged in a buckled honeycomb structure. Consequently, its behavior around the Fermi level is well described by the Kane-Mele Hamiltonian,
\begin{align}
    H &= t \sum_{\langle ij \rangle} \mathbf{c}^\dagger_i \mathbf{c}^{}_j + t_2 \sum_{\langle\langle ij \rangle\rangle} \mathbf{c}^\dagger_i \mathbf{c}^{}_j + t_3 \sum_{\langle \langle\langle ij \rangle\rangle \rangle} \mathbf{c}^\dagger_i \mathbf{c}^{}_j \notag \\ &+ \frac{i \lambda_{\text{SO}}}{3\sqrt{3}}\sum_{\langle\langle ij \rangle\rangle} \nu_{ij}\mathbf{c}^\dagger_i s_z \mathbf{c}^{}_j + i\lambda_\text{R} \sum_{\langle ij \rangle} \mathbf{c}^\dagger_i (\mathbf{s}\times\mathbf{d}_{ij}) \mathbf{c}^{}_j \notag \\ 
    &+ M_\text{S} \sum_i \eta_i \mathbf{c}^\dagger_i \mathbf{c}^{}_j ,
\end{align}
where $\mathbf{c}_i = (c_{i,\uparrow}, c_{i,\downarrow})^T$ is a vector composed of electronic annihilation operators, $t$ is the nearest-neighbor hopping parameter, $\lambda_{\text{SO}}$ is the SOC, $\lambda_\text{R}$ is the Rashba SOC and $M_\text{S}$ is the staggered (Semenoff) mass, $\nu_{ij}$ is $+1$ ($-1$) for (counter) clockwise hopping and $\eta_i$ is a staggering parameter, which is either $+1$ or $-1$, depending on the sublattice. Furthermore $\mathbf{s}=(s_x,s_y,s_z)$ are the Pauli matrices acting on spin space and d$_{ij}$ is the vector connecting sites $i$ and $j$. Additionally, longer range hopping terms, $t_2$ and $t_3$, have been included because first-principle calculations indicate that they are sizable in germanene.

In the absence of Rashba SOC, i.e. $\lambda_\text{R} = 0$, the two spin sectors decouple and each of them can be independently modeled by a Haldane Hamiltonian, 
\begin{align}
    H &= t \sum_{\langle ij \rangle} c^\dagger_i c^{}_j + t_2 \sum_{\langle\langle ij \rangle\rangle} c^\dagger_i c^{}_j + t_3 \sum_{\langle \langle\langle ij \rangle\rangle \rangle} c^\dagger_i c^{}_j \notag \\ &+ \frac{i \lambda_{\text{SO}}}{3\sqrt{3}}\sum_{\langle\langle ij \rangle\rangle} \nu_{ij} c^\dagger_i c^{}_j + M_\text{S} \sum_i \eta_i c^\dagger_i c^{}_j .
\end{align}
Here, we neglect the Rashba SOC.

\textit{Topological invariants. }The usual characterization of 2D topological systems is done in terms of (spin) Chern numbers. However, as a 2D system gets thinner, it can be modeled as periodic in the long direction and open in the direction perpendicular to the ribbon. Consequently, the spatial coordinate in the long direction can be Fourier transformed, resulting in a ribbon Hamiltonian. This Hamiltonian only has a single momentum index; the width of the ribbon gives rise to multiple bands instead of the two bulk bands of the usual Haldane model. Since this new ribbon Hamiltonian has a bulk that is parametrized by a single momentum index, its topological characterization is done in terms of a Zak phase. In particular, one must use the multiband Zak phase,
\begin{equation}
    \varphi = - \text{Im}\log\det \prod_{j=0}^{N-1} S(k_j, k_{j+1}).
\end{equation}
Here, $S(k_j, k_{j+1})$ is the overlap matrix of occupied Bloch functions, with elements $S_{mn} \equiv  
\langle u_{mk_j} | u_{nk_{j+1}}\rangle$. The Bloch eigenfunctions $| u_{nk_j} \rangle$ are obtained numerically by diagonalizing the ribbon Bloch Hamiltonian associated to Eq. (2) at (discretized) momentum $k_j=2\pi j/(Na)$, with $N$ and a denoting the lattice discretization number and lattice constant, respectively. A topological invariant $\nu$ is constructed by comparing the Zak phase to the one corresponding to the trivial limit, i.e. $M_\text{S} \to \pm \infty$,
\begin{equation}
    \nu \equiv \frac{\varphi-\varphi_{M_\text{S}\to\pm \infty}}{\pi} \mod 2.
\end{equation}
Further details regarding the calculations of the topological invariant $\nu$ for nanoribbons is presented in the SM.\\

\textit{Calculation of the LDOS. }Although the spectral properties of germanene nanoribbons are not experimentally accessible, the LDOS are promptly accessible by measuring the $dI/dV$. The LDOS at energy $E$ and position $r$ can be calculated using 
\begin{equation}
    \text{LDOS}(E,\mathbf{r}) = \frac{1}{\pi}\sum_n |\psi_n(\mathbf{r})|^2 \frac{b}{(E-E_n)^2 + b^2},
\end{equation}
where the wavefunctions $\psi_n(\mathbf{r})$ and the corresponding energies $E_n$ are obtained by numerically diagonalizing Eq. (1). Furthermore, an empirical broadening $b=0.05$ eV is incorporated to account for thermal effects, substrate scattering, and broadening due to lock-in techniques.

\end{document}